\documentclass{elsart}
\usepackage{graphicx}
\usepackage{amsmath}
\usepackage{amsfonts}
\usepackage{amssymb}
\begin{document}
\begin{frontmatter}
\title
{The Holographic Principle
and the Early Universe}
\author{F. Canfora and G. Vilasi}
\address
{Istituto Nazionale di Fisica Nucleare, Sezione di Napoli, GC di Salerno\\
Dipartimento di Fisica ''E.R.Caianiello'', Universit\`{a} di
Salerno\\ Via S.Allende, 84081 Baronissi (Salerno), Italy\\
e-mail: canfora@sa.infn.it}
\begin{abstract}
A scenario is proposed in which the matter-antimatter asymmetry behaves like a seed for the
inflationary phase of the universe. The mechanism which makes this scenario plausible is the
holographic principle: this scheme is supported by a good prediction of the number of e-folds.
It seems that such a mechanism can only work in the presence of a Hagedorn-like
phase transition. The issue of the "graceful exit" can be also naturally accounted for.
\end{abstract}
\end{frontmatter}

\section{Introduction}

\noindent The inflationary scenario \cite{Gu81} \cite{AS82} \cite{Li82}
\cite{Li83} has been one of the main achievement in theoretical cosmology of
the last decades. It provided many fundamental questions (such as why our
universe is flat, homogeneous and isotropic to a very high degree, why we do
not observe monopoles or other topological defects, why the primordial
perturbations have a flat spectrum and so on; detailed reviews are, for
example, \cite{Li90}, \cite{Gu97}) with a natural explanation. All the above
questions could be answered in a standard FRW model only assuming very special
initial conditions and fine tunings of many kinds. The mechanism which allows
to solve such problems is mainly based on a very fast initial expansion of the
scale factor of the universe which, in a sense, ''washes out'' the
inhomogeneities. The standard engine which drives such an expansion is a
scalar field, called \textit{inflaton}, which slowly rolls towards the minimum
of its potential. During the slow roll of the inflaton one gets a period of
exponential expansion of the universe. It is worth to stress here that, in the
inflationary scenario, what is really needed is the very fast initial
expansion (from which it is possible to deduce all the wanted physical
predictions by analyzing the evolution of the various kinds of perturbations),
the scalar field is the easier way to get it but there is no compelling
physical reason which tells that it is a scalar field, and not a vector or a
tensor field or a different kind of mechanism, to drive the inflation (see,
for example, \cite{Li90}, \cite{Gu97}). Actually, despite its striking
successes, the inflationary paradigm still has some problems which can be
traced back to the assumption that it is a scalar field to be responsible for
the inflationary phase. In particular, it is still not completely clear what
is the mechanism which allows a ''graceful exit'' from the inflationary phase,
there is not a commonly accepted potential for the inflaton, the physical
origin of the inflaton itself is still unknown and, \textit{a priori}, it is
not lawful to use the classical Einstein equations coupled to the inflaton
field, as it is usually done, to study the evolution of the universe in a
highly curved regime in which quantum corrections should be expected.

Besides the still unsolved problems of the inflationary scenario (which, on
the other hand, do not overshadow its great merits), theoretical cosmology is
affected by many unsolved problems. One of the most noticeable is the
matter-antimatter asymmetry (detailed reviews are, for example, \cite{KT90},
\cite{DK03}, \cite{BPY05}). At a first glance, the Lagrangian of the Standard
Model seems to be unable to explain why in the actual universe there is such
an amount of asymmetry between matter and antimatter which enter symmetrically
in the interactions. In a seminal paper \cite{Sa67}, Sakharov showed that this
asymmetry could be the consequence of the presence of baryon number violating
processes, CP violations and departure from thermal equilibrium. In fact, the
first two conditions can be fulfilled in the Standard Model: the effects are
very small but, in the early universe when the temperature was very high, they
are significantly enhanced and the third condition could also be met. There is
still not a commonly accepted explanation of this asymmetry; at the moment the
two most popular models seem to be the Leptogenesis (according to which the
weak interactions, converting some lepton number into baryon number, could
generate a net baryon and lepton number) and the Affleck-Dine mechanism based
on supersymmetry (according to which the scalar supersymmetric partners of
quarks and leptons could be responsible for the processes which should give
rise to a fulfillment of the Sakharov conditions).

Here, a scenario is proposed in which the matter-antimatter asymmetry is the
driving force of the inflationary phase of expansion of the universe. The
mechanism which makes this possible is the \textit{holographic principle} (up
to now, the most promising available open window on quantum gravity). From the
inflationary point of view, this mechanism also has the advantage of providing
a natural explanation of the ''graceful exit''.

In the first section we will briefly review the physical basis of the
holographic principle. In the second section we will describe a statistical
argument which, together with the holographic principle, makes plausible the
proposed scenario. Eventually, some conclusions will be drawn.

\section{The Holographic Principle}

Even if the quantum theory of the gravitational field has not been found yet,
in the few examples (such as the AdS/CFT correspondence \cite{Ma97}), in which
one can carry on quantum computations in the presence of a gravitational
field, the effective number of degrees of freedom is much smaller than the
number which one would naively expect on purely Quantum Field Theoretic
grounds: the number of degrees of freedom in a space-like region turns out to
be proportional to the area of that region. The holographic principle
heightens this phenomenon to a basic principle of the would be quantum theory
of gravity (see, for example, \cite{Bo02}); the physical basis of such a
principle were given in \cite{Be81}, \cite{Go83}, \cite{tH93} and \cite{Su95}.
Elegant refinements of the original ideas \cite{Be81}, \cite{Go83},
\cite{tH93} and \cite{Su95} can be found in \cite{Bo99}, \cite{BV99}. Even if
we still have not the final theory of quantum gravity, nevertheless it is
possible to argue that the holographic principle could have a prominent role
in understanding why the observed value of the cosmological constant is so
smaller than the one computed in Quantum Field Theory\footnote{of about 120
orders of magnitude in standard Quantum Field Theory which become 60 orders of
magnitude in supersymmetric QFT.} (henceforth QFT). An intuitive explanation
could be that in QFT pairs of degrees of freedom, which are coupled by
gravitational interaction to form ''bound states'' behaving as single
effective degrees of freedom, are counted as distinct. This overcounting could
be responsible of the too large cosmological constant obtained in QFT. In a
simple classical model \cite{CV05} it is also possible to find, without using
CFT, a direct purely holographic (although qualitative) relation between the
cosmological constant and the number of degrees of freedom%
\begin{equation}
\Lambda\sim\frac{\ln N}{N}\lambda\label{CCVV}%
\end{equation}
where $\Lambda$ is the cosmological constant, $N$ is (of the order of) the
total number of degrees of freedom of the universe and $\lambda$ is a
characteristic quantum energy density which cannot be determined in a
classical model but which, on dimensional analysis grounds, should be of the
order of the Planck mass to the fourth power. The above qualitative relation
between the cosmological constant and the number of degrees of freedom is in
good agreement with the so called \textit{N bound} \cite{Bo00} according to
which%
\begin{equation}
\Lambda\sim\frac{m_{P}^{4}}{N}. \label{sessea}%
\end{equation}

\section{Holography and the early universe}

Let us recall that, at least immediately after the end of inflation (when it
is usually placed the beginning of the baryogenesis \cite{KT90}), the universe
can be described with a very good approximation by its
statistical-thermodynamical properties. In particular, the local minima of the
free energy\footnote{It is known that gravity allows irreversible processes to
occur without ever reaching any unsurpassable maximum value of the entropy
\cite{To66}. However the evolution, in the very short time interval we are
considering, will be driven mainly by local maxima of the entropy.}
\begin{equation}
F=H-TS \label{free}%
\end{equation}
(where $H$ is the internal energy, $T$ the temperature and $S$ the entropy)
play a prominent role in determining the evolution of the universe during and
after the baryogenesis. Indeed, internal quantum processes have time scales
much smaller than the typical time scale of the cosmological evolution so that
there is enough time for the particles filling the universe to reach the
thermal equilibrium before the size of the universe changes significantly
(after all, the cosmological predictions based on this assumption are in good
agreement with observations). This statistical descriptions is inadequate
during the inflationary period which is likely to be an out of equilibrium
phase. However, a statistical description through the free energy should
provide with a detailed picture also immediately before the inflation. It is
usually assumed that, before the inflation smoothed out the inhomogeneities,
there had been a period in which all the different parts of the universe were
causally connected such that thermalization took place (\cite{Li90},
\cite{Gu97}): if this is the case, a description in terms of free energy
immediately before the inflation is certainly correct. Of course, in the
presence of a big-bang singularity this is not true. In fact, it is commonly
believed that the final theory of quantum gravity will resolve the initial
singularity: good proposals, for example, are available in loop quantum
gravity \cite{BJ01} and string theory \cite{GV03}. The question is: what
mechanism drives the universe out of equilibrium in the inflationary phase and
how such an out of equilibrium phase terminates? The following consideration
is useful. In the early universe the temperature was very high and in the free
energy (\ref{free}) the second term should had been dominating so that the
minimum of the free energy was determined by the maximum of the entropy.

Thus, before the inflation, the particles filling the universe should had
tended toward the maximum entropic state. Let us suppose that, at that time,
particles and antiparticles entered the microscopic interactions almost
symmetrically (as it happens in actual the standard model Lagrangian): this
implies that the number of particles should had been almost equal to the
number of antiparticles. Is this state with an equal number of particles and
antiparticles the maximum entropic states? The answer is no. Under certain
reasonable hypothesis, which will be described below, it is vastly more
countenanced a state in which there are only particles. The following argument
clarifies this point providing, at the same time, with a promising order of
magnitude estimate. Let us suppose that we have to set $N$ bosons and the
relative $N$ antibosons in the quantized energy levels of a certain system.
Let us divide the quantized energy levels $\varepsilon_{j}$ in groups labelled
by the index $j$\footnote{In other words, the difference between energies
belonging to the same group are assumed to be very small.}. Let us denote the
number of states of the $j- $th group with $G_{j}$ and the number of particles
(which need not to be necessarily interpreted as the ''standard''\ particles
of the standard model but simply as the degrees of freedom living, before the
inflation, in the universe) living in that group as $N_{j}$. The sum of the
entropy of the particles and the entropy of the antiparticles is a reasonable
estimates of the total entropy. As it is shown, for example, in \cite{LL86},
the non equilibrium entropy of the system will be
\begin{equation}
S_{N,\overline{N}}\sim2\sum_{j}^{x}G_{j}\left[  \left(  1+\overline{n}%
_{j}\right)  \ln\left(  1+\overline{n}_{j}\right)  -\overline{n}_{j}%
\ln\overline{n}_{j}\right]  , \label{entro1}%
\end{equation}
where $\overline{n}_{j}$ is the mean occupation number of the quantum
states\footnote{Of course, if we maximize the above expression (\ref{entro1})
for the entropy, taking into account the usual constraints on the total energy
and the total number of particles, we get the Bose-Einstein distribution. But,
before the inflation, the above assumptions on the total number of particles
and total energy could be too strong.}:
\begin{equation}
\overline{n}_{j}=\frac{N_{j}}{G_{j}}, \label{numeromedio}%
\end{equation}
and the upper limit $x$\ can be roughly estimated as follows%

\begin{equation}
\frac{1}{2}x^{2}\sim\sum_{j}^{x}\overline{n}_{j}%
=total\,\,number\;of\;particles, \label{rela3E}%
\end{equation}
In the case in which $N$ bosons and $N$ antibosons are present
\[
x\sim\left(  \sqrt{N}\right)  ^{\eta}%
\]
where $\eta$, a positive number of order $1$, takes into account the
uncertainty on procedure of the replacement of discrete \textit{sum} with an
\textit{integral}. Instead, when we have at our disposal $2N$ bosons
$x\sim\left(  \sqrt{2N}\right)  ^{\eta}$ and the entropy is given by
\begin{equation}
S_{2N}\sim\sum_{j}^{\left(  \sqrt{2N}\right)  ^{\eta}}G_{j}\left[  \left(
1+\overline{n}_{j}\right)  \ln\left(  1+\overline{n}_{j}\right)  -\overline
{n}_{j}\ln\overline{n}_{j}\right]  , \label{entro2}%
\end{equation}
so that%

\begin{equation}
\frac{S_{2N}}{S_{N,\overline{N}}}\sim\frac{\sum_{j=\left(  \sqrt{N}\right)
^{\eta}}^{\left(  \sqrt{2N}\right)  ^{\eta}}G_{j}\left[  \left(
1+\overline{n}_{j}\right)  \ln\left(  1+\overline{n}_{j}\right)  -\overline
{n}_{j}\ln\overline{n}_{j}\right]  }{\sum_{j}^{\left(  \sqrt{N}\right)
^{\eta}}G_{j}\left[  \left(  1+\overline{n}_{j}\right)  \ln\left(
1+\overline{n}_{j}\right)  -\overline{n}_{j}\ln\overline{n}_{j}\right]
}.\qquad\label{ratio}%
\end{equation}
In the absence of a quantum theory of gravity, to find the explicit
expressions of the $G_{j}$ and $x$ is a hopeless task; however we only need
the order of magnitude. We should determine how the $G_{j}$'s depend upon the
$\overline{n}_{j}$'s.

Some well known features of ''Hagedorn phenomenology'' in string theory and
(large $N_{C}$ SUSY) QCD (see, for example, \cite{Ha98}, \cite{LTW03}
\cite{Ar05} and references therein) suggest an interesting way toward this
goal: it is strongly believed that at the Hagedorn temperature there is a
phase transition toward a deconfining phase such that below the Hagedorn
temperature the free energy $F\sim T^{n}$ (in case, with logarithmic
corrections) and above one gets $F\sim T^{n+1}$, the entropy having a similar
behavior. Thus, at the Hagedorn temperature, the exponent of the entropy as a
function of the typical energy scale jumps in such a way that, above the
transition, the entropy increases faster with the energy scale. The typical
energy scale $\varepsilon_{2N}$, when there are $2N$ bosons, is higher than
the scale $\varepsilon_{N-\overline{N}}$\ when there are $N$ bosons and $N$
antibosons; thus, in the former case, we must impose that the entropy
increases faster: this will be the physical motivation behind our assumption
Eq. (\ref{hagedorn}). In many string inspired matrix models (which, for
example, describe the thermodynamic of a gas of D0-Branes) first order phase
transitions are also possible in which the discontinuity of the free energy is
of order $N^{2}$ (see, for example, \cite{Se04} and references therein) $N$
being (of the order of) the number of degrees of freedom: this kind of phase
transitions gives rise to Eq. (\ref{ratio2}) too. \textit{Viceversa}, one
could search for (sufficient) conditions in order to get Eq. (\ref{ratio2})
(which is at the basis of the main result of the paper Eq. (\ref{entro6})): a
good condition would be the existence of a Hagedorn-like (first or second
order) phase transition at the scale $\varepsilon_{N,\overline{N}}$. It is now
apparent an intriguing relation between Hagedorn phase transition, the
holographic principle and inflation.

The previous considerations lead us to assume that there exists a critical
value $j_{C}$ of the label beyond which the increase of the mean number of
particles in the group $j+1$ with respect to the group $j$ ( $j>j_{C}$) is
less than in the case $j<j_{C}$: to be concrete
\begin{equation}
j\lesssim\left(  \sqrt{N}\right)  ^{\eta}\Rightarrow G_{j}\sim\left(
\overline{n}_{j}\right)  ^{\alpha};\ j\gtrsim\left(  \sqrt{N}\right)  ^{\eta
}\Rightarrow G_{j}\sim\left(  \overline{n}_{j}\right)  ^{\alpha+1}
\label{hagedorn}%
\end{equation}
In this case the ratio in Eq. (\ref{ratio}) can be roughly estimated replacing
the sums with integrals:
\begin{equation}
\frac{S_{2N}}{S_{N,\overline{N}}}\sim\left(  \sqrt{2N}\right)  ^{\eta}%
\sim\left(  10^{60}\right)  ^{\eta}. \label{ratio2}%
\end{equation}

It is now quite clear that, for $2N\sim10^{120}$ (which should be a reasonable
estimates of the number of degrees of freedom in the early universe), the
evolution, driven by the entropy, should prefer to have only particles. The
observations distinctly tell us that there exists a mechanism to obtain only
particles (see, for example, \cite{KT90}) but now the holographic principle
comes into play.

According to the holographic principle, the entropy of the universe should
fulfil an equation of the following type
\begin{equation}
S=\alpha A(\partial V) \label{entro4}%
\end{equation}
where $\alpha$ is a suitable constant and $A(\partial V)$ is the two
dimensional area of the boundary of a (Cauchy) hypersurface of constant time.
If, before the inflation, the number of particles is almost equal to the
number of antiparticles because they enter almost symmetrically the
interactions, then, from Eq. (\ref{entro1}), it follows
\begin{equation}
A(\partial V)\sim S_{N,\overline{N}}. \label{entro5}%
\end{equation}
On the other hand, the free energy would drive the universe towards a state in
which there are almost only particles (in such a way to achieve a very huge
benefit in entropic terms). The holographic principle does not allow this gain
immediately: it allows the attainment of the minimum of the free energy only
if Eq. (\ref{entro4}) is fulfilled. The question is: how much time does the
holographic principle employ to boost the value of area in Eq. (\ref{entro5})
to the value which allows the attainment of the minimum of the free energy? At
this stage of knowledge of quantum gravity it is not possible to give a
definite answer to this question; however, since it is strongly believed that
the holographic principle is a quantum-gravitational effects, one can assume
that the duration of the inflation will be of ''Planck'' order
\begin{equation}
\tau_{HP}\lesssim10^{-30}s. \label{time}%
\end{equation}
Summarizing: in order to reach the minimum of the free energy without
violating the holographic principle, the universe should had expanded, in a
interval of time of the order (\ref{time}), from an area corresponding to an
entropy of the order (\ref{entro5}) to an area of the order (\ref{entro2}). To
support the hypothesis proposed here, we have to give an estimate of the
number of e-folds $N_{E}$ predicted in this scenario. This is an easy
computation: the number of e-folds $N_{E}$ is defined as follows
\[
N_{E}=\ln\left(  \frac{a_{F}}{a_{I}}\right)
\]
where $a_{I}$ and $a_{F}$ are the scale factor of the universe immediately
before and immediately after the inflation. According to the present model,
being the linear size of the universe $a$
\[
a\sim\sqrt{A(\partial V)},
\]
the number of e-folds should be of the order
\begin{equation}
N_{E}=\ln\left(  \sqrt{\frac{S_{2N}}{S_{N-\overline{N}}}}\right)  \sim
\eta30\ln10. \label{entro6}%
\end{equation}
This result is in very good agreement with the commonly accepted value for the
number of e-folds which is expected to lie between 60 and 70 (see, for
example, \cite{Li90}, \cite{Gu97}). To the authors knowledge, there are not
theoretical models in which the number of degrees of freedom of the universe
and the number of e-folds (which, in principle, can be determined
independently from the observations) can be related in a direct and effective
way as in the present scenario: these two numbers are not independent. This
result is quite robust: one would get similar estimates for $N_{E}$ (which
differ of no more than one order of magnitude) by changing slightly $\eta$ in
Eq. (\ref{ratio2}). It is worth to stress here that the above result
(\ref{entro6}) has been obtained without using the Einstein equations which,
at that time, were likely to be corrected by quantum gravitational effects.
Another highly non trivial benefit of this model is that the problem of the
''graceful exit'' simply has disappeared: when the area is such that it is
possible to attain the minimum of the free energy without violating the
holographic principle the inflationary phase terminates and the evolution is
again driven by the minima of the free energy. It also appears natural to
place the baryogenesis immediately after the end of the inflationary period
(as it is usually assumed \cite{KT90}, \cite{DK03}): the free energy would
like the baryogenesis ''as soon as possible'' because of the entropy gain; the
holographic principle prevents this but, when it has been fulfilled, the
baryogenesis can freely start. It is interesting to note that, according to
the results in \cite{Bo00} (further supported by the analysis in\ \cite{CV05})
it is possible to obtain a direct relation between the cosmological constant,
\textit{via} eqs.(\ref{sessea}) and (\ref{CCVV}), the number of degrees of
freedom and the number of e-folds which still is in good agreement with observations.

\section{Conclusion}

In this paper a scenario has been proposed in which the inflationary phase is
generated by the matter-antimatter asymmetry. The physical mechanism which
makes this possible is the holographic principle. The scheme is the following:
in the early universe the temperature was very high so that the entropy should
had driven the evolution toward its maximum. At that time, interactions in
which particles and antiparticles enter almost symmetrically are not able by
themselves to generate the observed asymmetry. On the other hand, a state in
which there are only particles is vastly more countenanced from the entropic
point of view thus, having at our disposal many microscopical mechanisms which
are able to generate the observed asymmetry at high temperatures, one could
expect that the evolution freely drives the universe toward the observed
asymmetric state. Now the holographic principle comes into play. According to
this principle, which is likely to be a manifestation of quantum gravity, the
entropy is tied to the area of the universe. For this reason, it is possible
to attain the minimum of the free energy only after that the holographic
principle boosted (in a Planckian time scale) the initial area of the universe
to a size compatible with the maximum of the entropy. The number of e-fold is
in a very good agreement with the commonly accepted one. Moreover, the problem
of the ''graceful exit'' from the inflationary phase has disappeared: the
inflationary phase terminates simply when the size of the universe allows the
attainment of the minimum of the free energy which is again allowed to drive
the evolution. This would also explain in a natural way the fact that, very
likely, the baryogenesis started immediately after the end of the inflation.
This scheme is likely to work only if one assumes the existence of a
Hagedorn-like phase transition. Even if more detailed computations are needed
to provide this proposal with further supports, we believe that its merits
make the physical basis of this model quite sound.

\end{document}